\documentclass[a4paper,fleqn,usenatbib,useAMS,usefloat ]{mnras}
\usepackage{mathptmx}
\usepackage[T1]{fontenc}
\usepackage{ae,aecompl}
\usepackage{graphicx}	% Including figure files
\usepackage{amsmath}	% Advanced maths commands
\usepackage{amssymb}	% Extra maths symbols
\usepackage{subfig}
\usepackage{ulem}
\usepackage{epstopdf}
\newcommand{\qmin}{\mathcal{Q}_{y}}
\newcommand{\qminS}{\mathcal{Q}_{y(S)}}

\title[Algorithm to derive the mass-ratio distribution of binaries]
{Study of the mass-ratio distribution of spectroscopic binaries. I. A novel algorithm}
\author[S. Shahaf et al.]{
S. Shahaf,\thanks{E-mail: sahar@wise.tau.ac.il}
T. Mazeh,
S. Faigler
\\
School of Physics and Astronomy, Tel Aviv University, Tel Aviv 69978, Israel
}

\date{Accepted 2017 August 30. Received 2017 August 29; in original form 2017 July 12}

\pubyear{2017}

\begin{document}
\label{firstpage}
\pagerange{\pageref{firstpage}--\pageref{lastpage}}
\maketitle

\begin{abstract}
%==============
We developed 
a novel {\it direct} algorithm to derive the mass-ratio distribution (MRD) of short-period binaries from an observed sample of single-lined spectroscopic binaries
(SB1). The algorithm considers a class
of parameterized MRDs and finds the set of parameters that best fits the observed sample.
The algorithm consists of four parts. First, we define a new observable, the `modified mass function', that can be calculated for each binary in the sample. We show that the distribution of the modified mass function follows the shape of the underlying MRD,  turning it more advantageous than the previously used mass function, reduced mass function or reduced mass function logarithm.
Second, we derive the likelihood of the  sample of modified mass functions to be observed given an assumed MRD.  An MCMC search enables the algorithm to find the parameters that best fit the observations.
Third, we  suggest to express the unknown MRD
by a linear combination of a basis of functions that spans the possible 
MRDs. We suggest two such bases.
Fourth, we show how to account for the undetected systems that have an RV amplitude below a certain threshold. 
Without the correction, this observational bias suppresses the derived MRD for low mass ratios. 
Numerous simulations show that the algorithm works well with either of the two suggested bases.
The four parts of the algorithm are independent, but the combination of the four turn the algorithm to be highly effective in deriving the MRD of the binary population.
\end{abstract}

\section{Introduction}
%================

The study of the mass-ratio distribution (MRD) of binaries, short-period ones in particular, has a long
history. This is so because the MRD plays a key role in various aspects of the theory of binary formation
 and evolution. First, it provides one of the very few ways to
confront the theories of binary formation   \citep[e.g.,][]{bate97,satsuka17} with
 observations.
Second, the primordial MRD is a one of the few major inputs for population syntheses of binaries \citep[e.g.,][]{toonen12,yungelson17}, which try to predict, for example, the rate of supernova explosion and black hole mergers.
Third, the binary fraction and the MRD have been shown to play an important role in star cluster evolution \citep[e.g,][]{hut92,benacquista13}.
Finally, understanding the low end of the MRD is crucial for the determination of the borders of the brown dwarf desert \citep[e.g.,][]{mazeh03,grether06} that separates exoplanets from low-mass stellar secondaries. 

It is therefore not surprising that quite a few studies tried to derive the MRD of binaries, using short-period spectroscopic binaries (SB) in particular. Reviews of the early studies  can be found in \cite{trimble90} and \cite{mg92}.
Because some of these studies  
\cite[e.g.,][]{lucy79,DM91,tokovinin00,goldberg03,fisher05,raghavan10,boffin10,boffin15,cure15}
%--------------------------------------------------------------------------------------------------------------------
%
yielded conflicting results, the shape of the MRD of short-period binaries continues to be an open question.

Derivations of the MRD should be based on a complete sample of binaries discovered by a systematic radial-velocity (RV) search. 
Generally, MRDs may depend on the mass of the primary star \cite[][]{kouwenhoven09}, therefore analyzed samples should be restricted to some narrow range of spectral types.
In an ideal world, with spectra of unlimited
resolution and signal-to-noise ratio, each SB would be a double-lined
binary (SB2), with mass ratio derived directly from the ratio of the RV amplitudes of the two components. 
In reality, 
the derivation of the MRD of short-period binaries is based on samples for which most of the binaries observed are single-lined spectroscopic binaries (SB1), where only the RVs of the primary star can be measured. Even after obtaining observations of a large sample of SB1s, deriving the MRD is hampered by the fact that for each of the SB1s the orbital solution cannot yield the mass ratio itself but merely
the mass function, a combination of two unknowns---the mass ratio and plane-of-motion
inclination angle.
Therefore, a statistical approach must be applied to the observational results, assuming {\it random distribution of the orbital inclination} of the sample as a whole. 

Two main approaches have been used to disentangle
the MRD from the orbital inclination \citep[see, for example,][]{heacox95}. In the inverse approach, one considers the sample of derived mass functions and 
works his/her way back, usually iteratively \citep[see the classical work of][]{lucy74}, to the underlying MRD of the binary population \citep[e.g.,][]{lucy79,mg92,boffin10,cure15}. 
%--------------------------------------------------------------------------

In the direct approach,
on the other hand,  one assumes a certain MRD of the binary population, calculates the resulting {\it expected} distribution of some observable, $O$, 
and compares it with the set of \{$O_i$\} obtained from the sample of SB1s, where the $i$-th binary is represented by $O_i$. 
One then finds the MRD that best fits the observed set \{$O_i$\} \citep[e.g.,][]{tokovinin92,hogeveen92,ducati2011}. 
The comparison between the expected distribution and the observed sample is usually  done by comparing histograms, not a very powerful approach, which does not allow statistical derivation of the best parameter(s) of the MRD and its (their) confidence intervals.

In previous studies, the observable used was the reduced mass function, $y$, 
obtained by dividing the mass function by the mass of the primary star, but see \cite{lucy79,boffin10,boffin15}, 
who promoted the use of a logarithm of the observable instead. 
However, we will show that the expected distributions of both $y$ and $\log y$ for very different MRDs are quite similar, respectively, turning the derivation of the true MRD quite difficult.

Here we present a novel algorithm to solve the problem, which consists of four parts. First, we introduce a new observable, $S$, which we coin the `modified mass function', that is derived 
for each SB1. 
We then show  that the shape of the distribution of the obtained $S$ for a binary sample is similar to that of the underlying MRD of the population. Even more importantly, different MRDs result in different $S$-distributions.

Second, we suggest to compare the expected distribution of $S$ with the observed sample by deriving the {\it likelihood} of the observed set of $\{S_i\}$, given the assumed MRD \citep[see also][]{tokovinin92}.
We can then find the parameters of the MRD that maximize the likelihood of observing the sample, using MCMC approach, for example.

Third, we  suggest to express the unknown MRD
by a linear combination of a basis of functions, with some unknown coefficients. 
We then search for the coefficients that  maximizes the likelihood of the observed sample.

Fourth, following \cite{mg92} we show how to account for the undetected binaries that have an RV amplitude below a certain threshold. 
Without the correction, this observational bias suppresses the derived MRD for low mass ratios. 

Section 2 introduces the modified mass function, Section 3 details the search for the best set of parameters using MCMC process, and suggests two sets of functions, and Section 4 brings two simulated examples which demonstrate that the algorithm works well. Section 5 presents our correction function, and Section 6 briefly summaries this work and lays out  possible further refinements of the algorithm.
In the next papers we apply the algorithm to various SB1 samples. 

\section{The modified mass function}
%==========================

\begin{figure*}
	\centerline{\includegraphics[width=0.75\linewidth,trim={0 0 0 0 },clip]{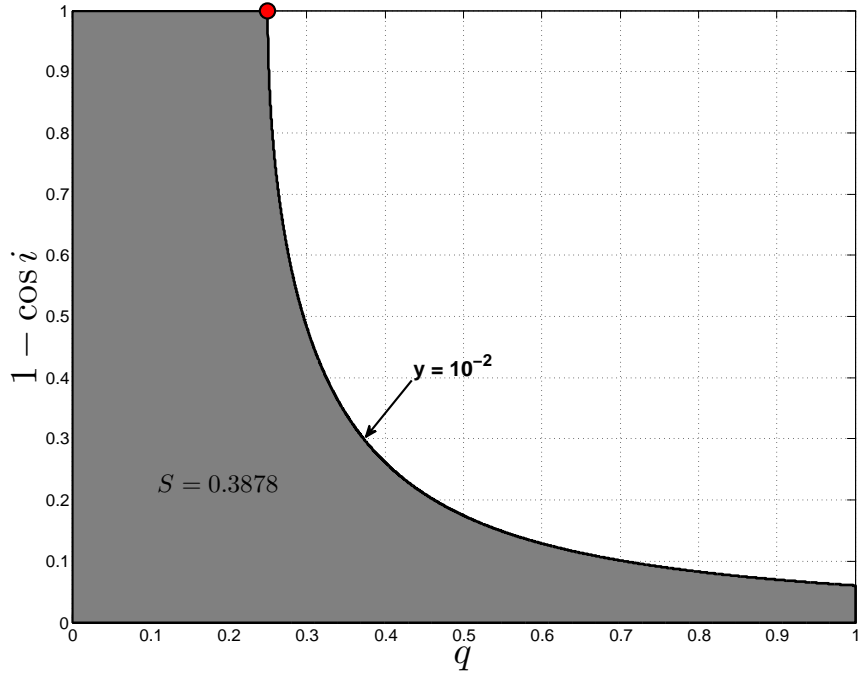}}
	\caption{The $(1-\cos i,q)$ parameter space, with an SB1 with $y=10^{-2}$. Gray area---the corresponding $S$ value. Red dot marks the minimum value of $q$---$\qmin$, of 0.25. }
	\label{fig:S_plot}
\end{figure*}

The mass-ratio of a binary system is defined as $q\equiv m_2/m_1 $,
where $m_1$, $m_2$ are the stellar masses of the primary and secondary, respectively.
For an SB1, only the spectrum of $m_1$ is seen in the spectrum of the system, and therefore only the primary RVs are obtained.  
When enough measurements are secured an orbital solution is derived, yielding the orbital period, $P$, eccentricity, $e$, 
and primary RV semi-amplitude, $K_1$.
The binary mass function is defined as
\begin{equation} \label{EQ: mass function}
	f(m_1) \equiv \frac{P\,K_1^3}{2\pi\,G}\,\big (1-e^2\big )^{3/2}=  m_1\,\frac{q^3}{(1+q)^2} \,  sin^3i \,,
\end{equation}
where $i$ is the orbital inclination.
An estimate of $m_1$ is often available from the binary spectra and can be factored out, leaving a `reduced mass function', $y$, with only two unknown parameters,  $q$ and $i$:
\begin{equation} \label{EQ: norm mass function}
	y \equiv \frac{f(m_1)}{m_1} = \frac{q^3}{(1+q)^2} \,  sin^3i \,.
\end{equation}
Henceforth, we assume that the primary is also the more massive star in the binary system, namely $0<q\leq1$. Under this assumption the reduced mass function becomes bounded as well, $0< y \leq 0.25$.
%
%, where practically only positive values of $y$ are detectable.
%
The relation between $q$, $y$ and $i$  is plotted in Figure~\ref{fig:S_plot}.
We choose to work in the $(1-\cos i, q)$ plane, as the distribution of $1-\cos i$ is uniform for random orientation of the orbits. 
The plot shows the possible values of $q$ for a binary with $y=10^{-2}$.
The gray area in the plot shows all the possible cases with $y\leq 10^{-2}$.

%-----------------------------------------------------------------------------
%-------------------------------------------------------------------------------
%

Notably, each value of $y$ is associated with a minimal possible $q$ value, that can be determined by setting the inclination angle $i$ in equation~(\ref{EQ: norm mass function}) to be $90^{\circ}$. 
This $q$ minimum, $\qmin$, is the only real root of the polynomial $P_{y}(q)$,
\begin{equation} \label{EQ: qmin poly}
		P_{y}(\qmin)=y^{-1}\qmin^3 -\qmin^2 - 2\qmin -1 = 0\,,
\end{equation}
for which an explicit expression was given by \citet{heacox95} (see also equation~(\ref{EQ: qmin})).  
That value, $0.25$ for $y=10^{-2}$, is plotted as a red point in the diagram.

Previous techniques used the observable $y$ or
$\log y$ as tools for deriving the MRD. For the direct method one needs to obtain the probability density function (PDF) of $y$, $f_y$, or $f_{\log y}$, given the PDF of the MRD, $f_q$. 
To obtain $f_y(y \, ; \, f_q)$ one has to calculate the probability to get a value between $y$ and $y+dy$ over the parameter space of Figure~1, given $f_q(q)$. This is done in 
Appendix \ref{APP: reduced mass function distribution}.

We seek a transformation $S=\mathbb{S}(y)$ such that significant functional properties of the MRD, $f_q$, will be qualitatively demonstrated by its resulting PDF, $f_S$. 
This results in three requirements.   
First, a uniform $f_q$ should yield a uniform $f_S$. 
Second, the transformation $\mathbb{S}$ is required to be strictly increasing and continuous.
Finally, for $f_S$ to be comparable with $f_q$, the range of $\mathbb{S}$ is required to be the $[0,1]$ interval. 

\begin{figure*}
	\centerline{\includegraphics[width=0.75\linewidth,trim={0 0 0 0 },clip]{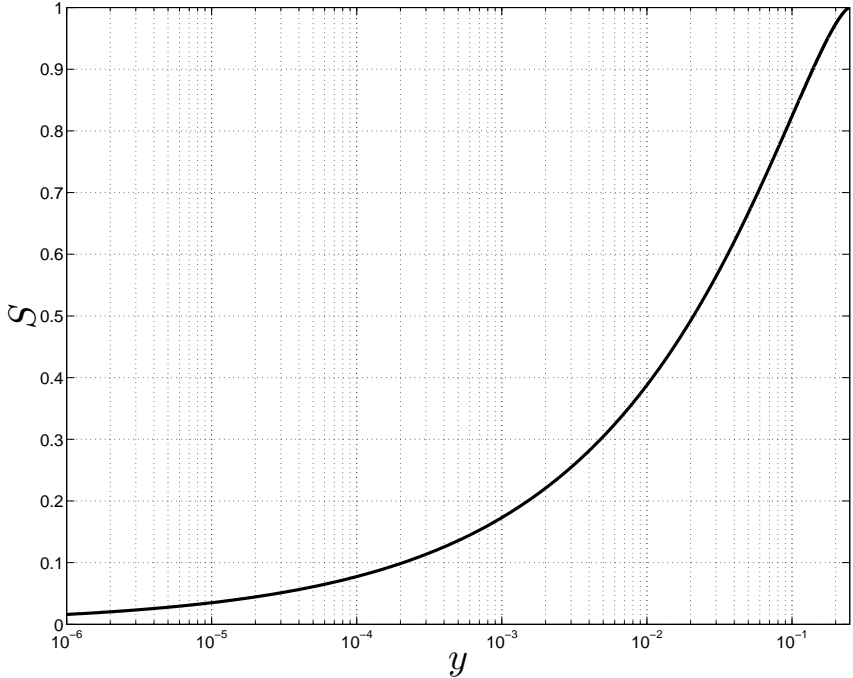}}
	\caption{Modified mass function $S$ as a function of the reduced mass function $y$.}
	\label{fig:S_vs_y}
\end{figure*}

These requirements are uniquely met by the cumulative distribution function (CDF) of $y$, {\it assuming a uniform distribution of $q$}.
For example, $\mathbb{S}(y=10^{-2})$ is simply the gray area of 
Figure \ref{fig:S_plot} for $y=10^{-2}$. 
The area can be written as  the integral
%
%---------------------------------------------------------------
\begin{equation} \label{EQ:S}
S=\mathbb{S}(y) \equiv 1 - \int_{\qmin}^{1}{\sqrt{1 - y^{2/3} \,{(1+q)^{4/3}}{q^{-2}}}\,dq} \,,
\end{equation}
%------------------------------------------------------------
%
where the integrand is the height above the curve of
Figure \ref{fig:S_plot}.
%The definite integral, $S$, hereafter named the `modified mass function'. 
%
The relation between $S$, hereafter named the `modified mass function', and $y$ is demonstrated in Figure \ref{fig:S_vs_y}.

The modified mass function, by its definition, resembles to a copula \citep[e.g.,][]{nelsen13}. While copulas are widely used in many fields, especially in quantitative finance, its astrophysical applications are rare \cite[for example]{scherrer10}.
Detailed derivation of  $\mathbb{S}$ appears in Appendix~\ref{APP: S derivation}.

%------------------------------------------------------------------------------
%-------------------------------------------------------------------

\subsection{Distribution of the modified mass function}
%--------------------------------------------------------------------

To derive the PDF of $S$ we note that $S$ is defined as a function of $y$, and therefore 
%------------------------------------------------
\begin{equation} 
f_S(S \,\, ; \, \, f_q) = 
f_{y}\big(y(S) \,\, ; \,\, f_q \big) \cdot \bigg|\frac{dy}{dS} \bigg| \,,
\end{equation}
%------------------------------------------------
%
where $y(S)$ is the inverse of $\mathbb{S}$. 
Since the modified mass function, $S$, is by definition the CDF of $y$ for a uniform MRD, the last factor in the equation above is 
%------------------------------------------------
\begin{equation} 
 \bigg|\frac{dy}{dS} \bigg| 
=
\frac{1} 
{ f_{y}\big(y(S) \,\, ; \,\,  1 \big)} \,.
\label{EQ: dS/dy}
\end{equation}
%------------------------------------------------
%
We therefore get
%
%------------------------------------------------
\begin{equation} 
f_S(S \,\, ; \, \, f_q) = 
f_{y}\big(y(S) \,\, ; \,\, f_q \big) \cdot \bigg|\frac{dy}{dS} \bigg| =
\frac{f_{y}\big(y(S) \,\, ; \,\, f_q \big)} { f_{y}\big(y(S) \,\, ; \,\,  1 \big)} \,.
\label{EQ: f_S in terms of f_y}
\end{equation}
%------------------------------------------------

An explicit expression for $f_y$ is developed in Appendix \ref{APP: reduced mass function distribution}, and shown in equation~(\ref{EQ: f_y APP}). 
Inserting the two expressions---the PDF of $y$ for the actual MRD and for flat distribution, we finally get
%
%---------------------------------------------------------
\begin{equation}
f_S(S \,\, ; \, \, f_q) = {\int_{\qminS}^1 {f_q(q) \, \mathbb{K}(y(S),q) \, dq }} \bigg/  
{\int_{\qminS}^1 {\mathbb{K}(y(S),q) \, dq }}  \,,
\label{EQ:f_S}
\end{equation}
%--------------------------------------------------------------
where 
%
%---------------------------------------------------------
\begin{equation}
\mathbb{K}(y,q) 
= \frac{(1+q)^{4/3}}{ 3 \, y^{1/3} \, q \,  \sqrt{ q^2 - y^{2/3}(1+q)^{4/3} }  }  \,,
\end{equation}	
%---------------------------------------------------------------
%
(see  Appendix \ref{APP: reduced mass function distribution}).
Equation~(\ref{EQ:f_S}) is effectively a weighted average of $f_q$ for a given $S$ value, taken over the allowed $q$ range, $[\qminS,1]$, 
and weighted by the assumed distribution of isotropic inclination angles.\\

\begin{figure*}
	\centering
	\includegraphics[width=1\linewidth ,trim={0cm 0cm 0cm 0cm },clip]{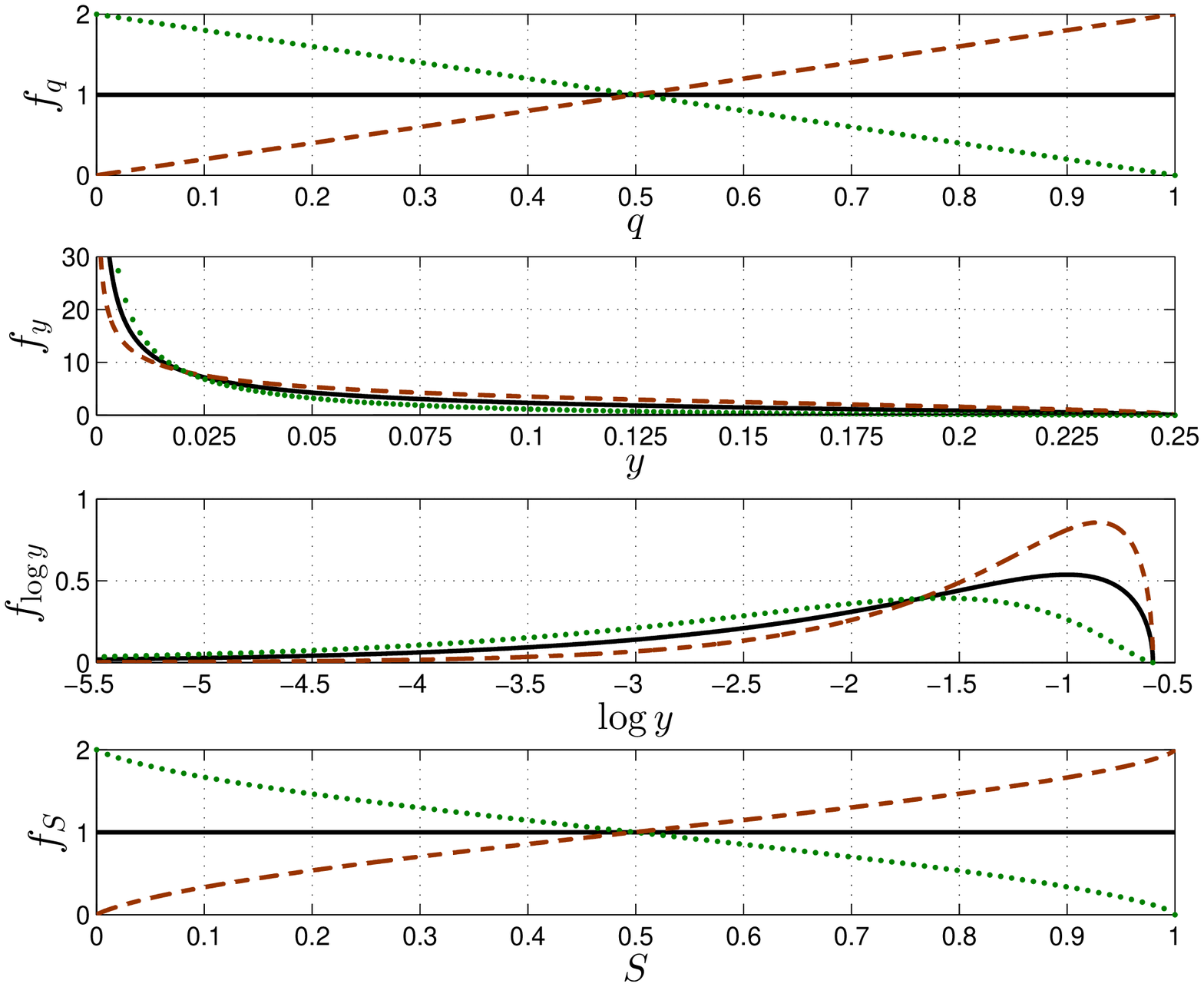}
	\caption{
Three distributions of $q$ and their corresponding $y$, $\log y$ and $S$ distributions. 
	Top panel: Uniform (black line), linearly increasing (brown dashed line) and decreasing (green dotted line) distributions of $q$.
The three lower panels show the  corresponding $y$, $\log y$ and $S$ distributions, with the same color and shape lines. 
}
	\label{fig:S_PDF}
\end{figure*}

Figure \ref{fig:S_PDF} shows the derived $f_y$ and $f_S$ for three different simple $f_q$ functions, demonstrating how, unlike $f_y$ or $f_{\log y}$, $f_S$ captures the shape of the 
underlying MRD, $f_q$.

%---------------------------------------------------------------
\section{Direct derivation of the MRD}
% ==================================== 
% 

\subsection{Likelihood derivation of the MRD}
% --------------------------------------------------------
\label{subsection: likelihood}
Let us assume that the MRD is characterized by a set of parameters ${\bf{c}}=\{c_k\}$. This could be, for example, 
a Gaussian distribution $f_q\propto \exp \big( (q-c_1)^2/2c_2^2 \big)$,  
a power-law distribution $f_q\propto q^{c_1}$,  a flat distribution between $c_1=q_{\rm min}$ and $c_2=q_{\rm max}$, and alike, or any combination of the above. 
Using equation~(\ref{EQ:f_S}), we transform the $f_q(q;{\bf{c}})$ into  $f_S(S;{\bf{c}})$, which has the same set of parameters $\{c_k\}$.

We wish to find the values of the $c_k$'s that best match the given sample of observed SB1, with a set of  $\{y_i\}$  that we transfer via equation~(\ref{EQ:S}) to the  corresponding set of $\{S_i\}$.
The search is done by maximizing the log-likelihood of ${\bf{c}}$,
\begin{equation}
log\mathcal{L} ({\bf{c}} | \{S_i\}) \,,
\label{EQ: logL1}
\end{equation}
given $\{S_i\}$. The core of the algorithm is the search for the best MRD in the $S$ domain. 
Since the fitted $c_k$ values are shared by both $f_S$ and $f_q$, it is clear that by fitting $f_S$ one readily derives its underlying $f_q$.

In practice, examples brought in this work were analyzed with the emcee ensemble sampler \citep{goodman10,foreman-Mackey2013}.
Each step in the generated chain yields a set of values for  the $c_k$'s, from which the MRD, $f_q$, is derived over a dense set of pre-determined $\{q_j\}$.
The chain yields {\it a posteriori} distributions for each $\{f_q(q_j)\}$.
We use these distributions  to derive the median
$\{\hat{f}_q(q_j)\}$ and their $1\sigma$ confidence intervals $\{\hat{\delta}(q_j)\}$ to finally yield 
%
%------------------------------------------
\begin{equation}
 %\hat{f_q}(q) = \sum_k \hat{c_k} \,\, \phi_k(q) \pm \sum_k \sigma_k \,\, \phi_k(q) \,. 
 \hat{f_q}(q) \pm \hat{\delta}(q) \,. 
 \label{EQ:f_q}
\end{equation}
%---------------------------------------------------------

\subsection{Expansion of MRD by a set of basis  functions}
%-----------------------------------------------------------------------
%
Likelihood derivation of the MRD requires a predetermined functional model, whose parameters are searched to fit best the observed  set of $\{S_i\}$. 
It is therefore desirable to use models that can span a broad class of functions, thus avoiding {\it a priori} assumptions about the functional shape of the MRD. This can be achieved by
 approximating  $f_q$ with a set of basis functions,
\begin{equation}
 f_q(q) = \sum_k c_k \,\, \phi_k(q) \,, 
 \label{EQ: f_q expansion}
\end{equation}
where $\phi_k(q)$ is the $k$-th function and $c_k$ is its corresponding coefficient. 

For each basis function we derive its corresponding function in the $S$ plane,  
 denoted $\tilde{\phi}_k(S)$, through equation~(\ref{EQ:f_S}),
\begin{equation}
\tilde{\phi}_k(S) \equiv f_S(S \,\, ; \, \,  \phi_k) \,.
\end{equation}
The linearity of equation~(\ref{EQ:f_S}) links the expansion of $f_S$ to that of $f_q$ via the modified functions, namely
\begin{equation}
 f_S(S \,\, ; \, \, f_q) = \sum_k c_k \,\, \tilde{\phi}_k(S) \,,
 \label{EQ: f_S expansion}
\end{equation}
where $c_k$ is the $k$-th coefficient of the $f_q$ series expansion.
In this case, the parameterized probability density takes a simple form 
\begin{equation}
f_S(S_i | {\bf{c}}) = \sum_k \tilde{\phi}_k(S_i) \cdot c_k \equiv \sum_k {M}_{ik} \,\, c_k  \,,
\end{equation}
where  $M_{ik}$ is the value of $\tilde{\phi}_k$ at $S_i$. This can be written in a matrix form
\begin{equation}
{\bf{f_S}}(S_i | {\bf{c}}) = {\bf\rm M}\cdot {\bf{c}}\,,
\end{equation}
where ${\bf\rm M}$ is the $M_{ik}$ 
design matrix.

The design matrix ${\bf\rm M}$ can be calculated given the sample $\{S_i\}$, the basis $\{\phi_k(q)\}$ and its corresponding $\{\tilde{\phi}_k\}$.
The log-likelihood in terms of ${\bf\rm M }$ is
%
%-----------------------------------------
\begin{equation}
%log\mathcal{L} ({\bf{c}} | \{S_i\})= 
log\mathcal{L} ( {\bf{c}} \,|\,\{S_i\})= 
\sum_i {\rm log} \bigg(
\sum_k {M}_{ik} \,\, c_k \bigg) \,,
\label{EQ: logL}
\end{equation}
%--------------------------------------------
%
according to which the best ${\bf{c}}$ can be found. 

 In the next subsections we suggest two possible sets of basis functions---the shifted Legendre polynomials and the boxcar functions. The two have complementary properties in terms of smoothness and locality.
Obviously, other possibilities, such as harmonic functions or power series, can be considered and implemented in a similar manner. 

\subsubsection{Shifted Legendre polynomials}
%------------------------------------------------------------ 

\begin{figure*}
	\centering
	\centerline{\includegraphics[width=1\linewidth]{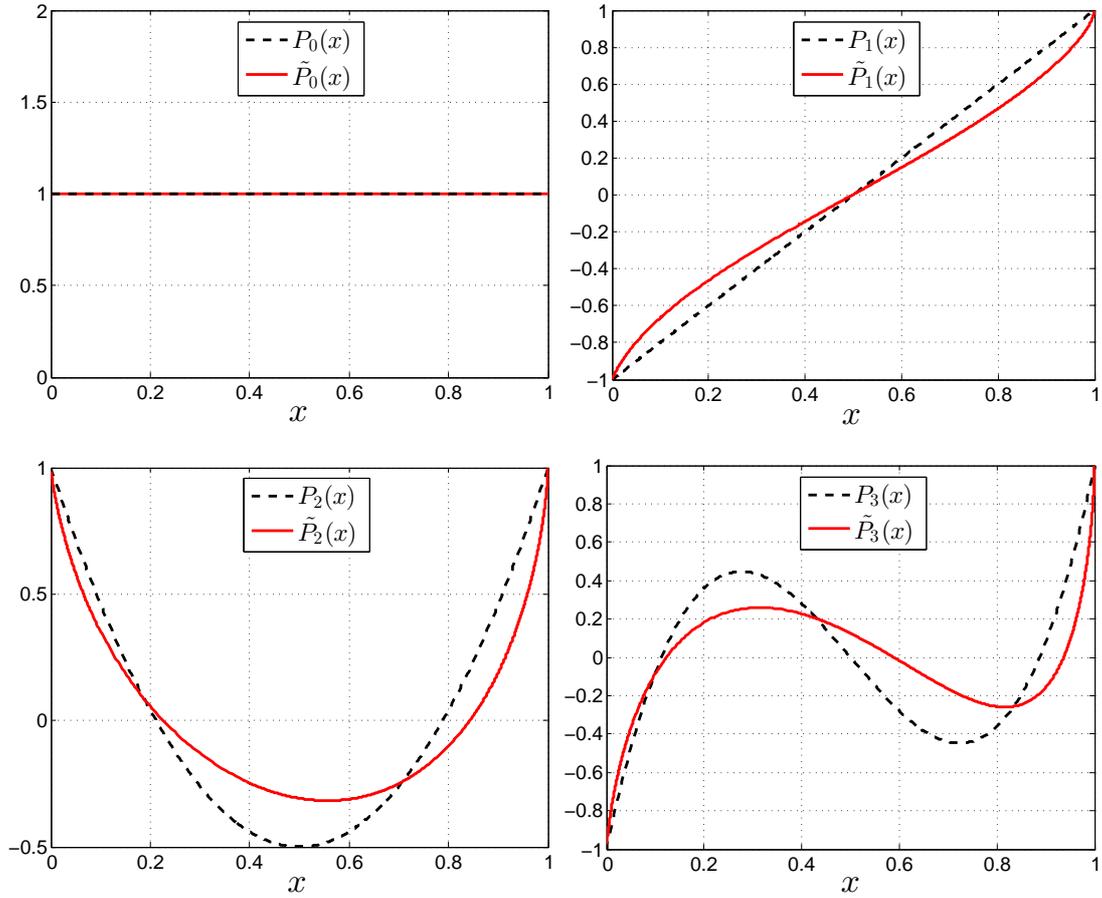}}
	\caption{First four shifted Legendre polynomials (dashed black) and their corresponding modified functions  (solid red).
}
	\label{fig: S legendre}
\end{figure*}

A possible basis is, for example, the shifted Legendre polynomials, $\{P_k\}$:
\begin{eqnarray}
\label{EQ: Legendre Poly}
P_0(x) &=& 1\,,\\
P_1(x) &=& 2x-1\,,\nonumber\\
P_2(x) &=& 6x^2-6x+1\,, \nonumber\\
P_3(x) &=& 20x^3 - 30x^2 + 12x -1\,,\nonumber\\
P_4(x) &=& 70x^4 - 140x^3 + 90x^2 - 20x + 1\,.\nonumber\\
      &\vdots& \nonumber \
\end{eqnarray}
The first four $P_k(q)$, $P_0$ -- $P_3$, are plotted in Figure~\ref{fig: S legendre},  together with their corresponding modified functions $\tilde{P}_k(S)$.

The shifted Legendre polynomials have the property 
$\int_0^1 P_k(x)dx = \delta_{k0}$. 
This makes them suitable as a set of basis function for any PDF, as the integral of any combination of them over the range [0,1] is unity, as long as $c_0=1$.
%
%---------------------------------------------------------------------------------
%---------------------------------------------------------------------------------
%

\subsubsection{Boxcar functions}
%--------------------------------------

Another basis  is the set of boxcar functions, $\{\Pi_{_{N,k}}(x), k=1,...,N\}$, that are simple unit pulses of the form
%
%----------------------------------------------------------------------
\begin{equation}
\Pi_{_{N,k}}(x) =
    \begin{cases}
   1  & \text{if }  \ \frac{k-1}{N} \leq x \leq \frac{k}{N} \,,
    \\
    0 & \text{else}\,,
    \end{cases}
\label{EQ: boxcar}
\end{equation}
%---------------------------------------------------------------
spanning the histogram-like models. 
%For such a model to be a PDF of $f_q$, the sum of the coefficients must be equal to unity. 
The corresponding modified functions are derived through equation~(\ref{EQ:f_S}).

\subsection{Starting point of the MCMC}
 In any MCMC search, it is important to start the chain not too far from the global maximum of the log-likelihood. Our starting point relies on the 
histogram of observed $\{S_i\}$. For the boxcar basis the starting point is taken as the normalized number of counts in the $S$ histogram bins,
whereas the starting point of the Legendre polynomial set was derived by a simple 
 linear least squares fit to the histogram bins  \citep[see][]{barlow89}.

To choose the number of bins, $N_{bin}$, for the histogram we use the Rice rule \citep{terrell85}, 
%-----------------------------------------------------------
\begin{equation}
 N_{bin} = \big\lceil \, \sqrt[3]{2n} \,\, \big\rceil \,,
\end{equation}
%-----------------------------------------------------------
where $n$ is the size of the observed sample.

\section{Testing the algorithm}
%========================

In order to test our algorithm, and the two bases presented above in particular, we ran numerous simulations, two of which are presented here. 
In each numerical experiment we assumed an underlying MRD and prepared a simulated SB1 sample, drawing at random values for the mass ratio and inclination of each binary. 
We then derived the $S$ value for each binary and applied our algorithm to the sample of modified mass functions twice, using in each time one of our two bases. 

{\it In all our simulations we were able to retrieve the correct shape of the underlying MRD}, with each of the two bases.
 
\begin{figure*}
	\centering
	\includegraphics[width=15cm, height=10cm]{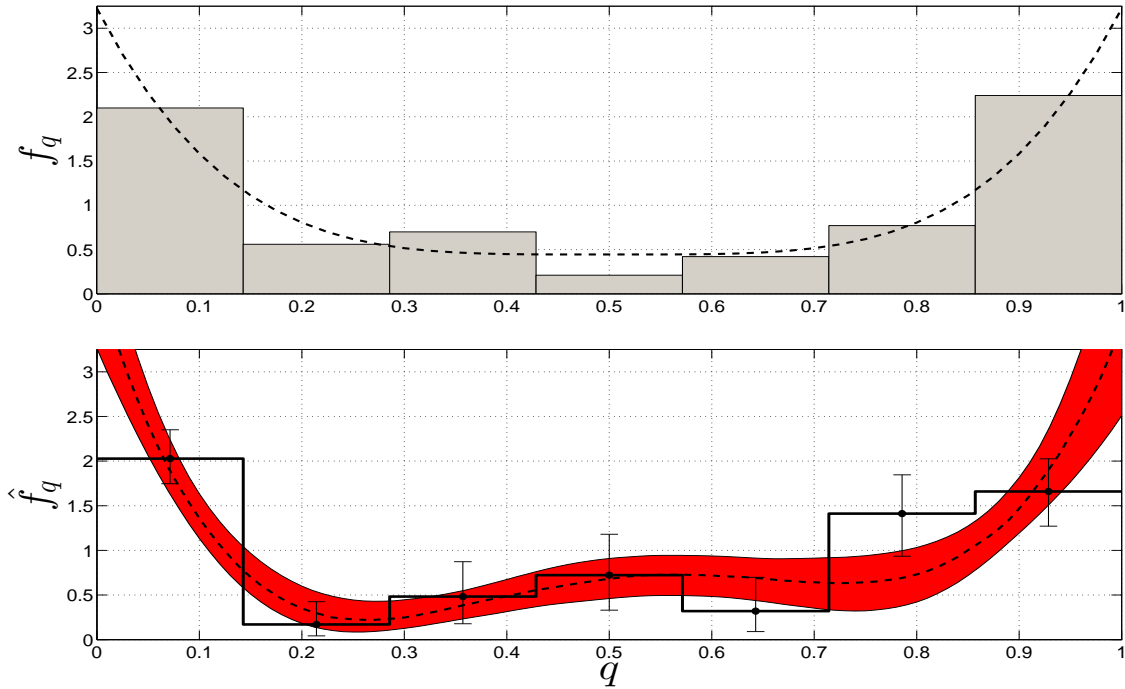}
	\caption{Derivation of MRD from a simulation 100 SB1 sample.
Top: Simulated sample of 100 SB1s, with random orientations, using as an MRD (dashed black line) 
a fourth-degree polynomial, 
 $f_q(q) \propto 25(2q-1)^4 + 4$, 
that peaks at $q=0$ and $q=1$. The specific drawn sample is presented by a seven-bin histogram.
Bottom: Two  independent derived MRDs, one uses the first {\it five} shifted Legendre polynomials as a basis (dashed black line) and the other one the boxcar basis of seven bins (thick black line). Each derived MRD is associated with an error for each value of $q$ (see text).
}
	\label{fig:Simulation1_fit}
\end{figure*}

\begin{figure*}
	\centering
	\includegraphics[width=15cm, height=10cm]{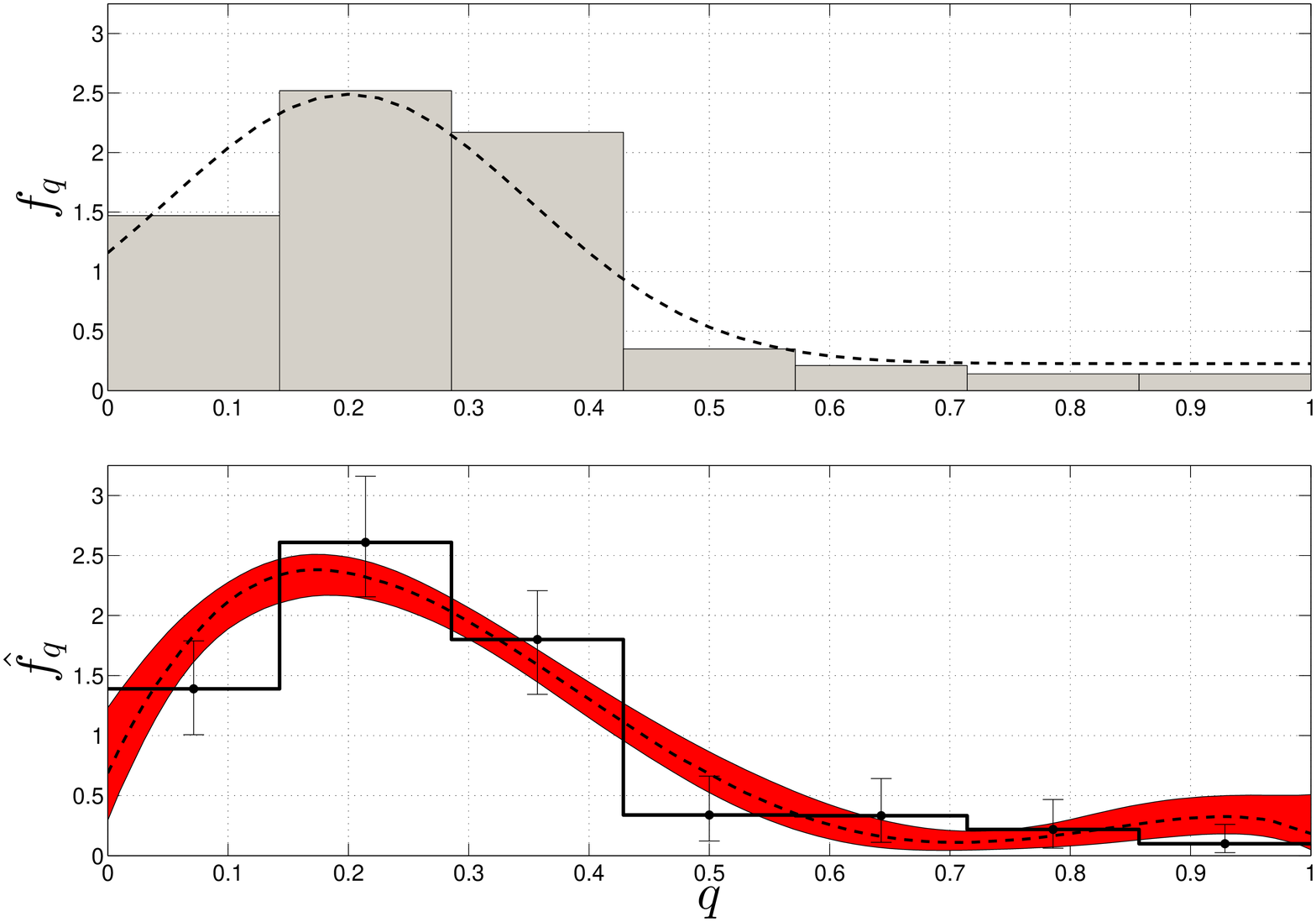}
	\caption{
Derivation of MRD from a simulation 100 SB1 sample.
The simulated MRD is composed of a Gaussian with a mean at $q=0.2$ and width of $0.15$ (77\% of the population) and a flat part in the range $q=[0,1]$ (23\%). 
The simulated sample and the two derived MRDs are plotted as in Figure~\ref{fig:Simulation1_fit}.
}
	\label{fig:Simulation2_fit}
\end{figure*}

Here we present two simulations, one  (Figure~\ref{fig:Simulation1_fit}) with an MRD composed of a fourth-degree polynomial, 
 $f_q(q) \propto 25(2q-1)^4 + 4$, 
that peaks at $q=0$ and $q=1$, and the other 
(Figure~\ref{fig:Simulation2_fit}) composed of a Gaussian with a mean at $q=0.2$ and a standard deviation of $0.15$ (77\% of the population) together with a flat distribution in the range $q=[0,1]$ (23\%). 
Since typically the number of SBs in modern spectroscopic surveys is on the order of 100 \citep[analyzed 129 SBs, for example]{goldberg03}, we chose the size of the simulated sample to be 100 SB1 systems in both examples.

The best MRD model and its uncertainty were derived by calculating the median and scatter of the values obtained for each $q$ along the MCMC run, as described in subsection~\ref{subsection: likelihood}.
The top panels of Figures~\ref{fig:Simulation1_fit} and~\ref{fig:Simulation2_fit} show the MRD used and the mass-ratio histogram of the simulated sample, 
while the bottom panels present the MRDs derived with a basis of seven boxcar functions and with the first five shifted Legendre polynomials.

An alternative method of deriving an explicit expression for the best fitting model is by taking median value of each parameter obtained along the chain. 
For example, the fitted  MRD, 
in terms of the shifted Legendre polynomials given in equation~(\ref{EQ: Legendre Poly}), for the two experiments presented above in this section are
\begin{align*}
&\hat{f}_1(q) =   1.22\,P_4(q) - 0.35\,P_3(q) + 1.54\,P_2(q) + 0.11\,P_1(q) + P_0(q)\,,\\
&\hat{f}_2(q) =   - 0.65\,P_4(q) + 0.99\,P_3(q) + 0.14\,P_2(q) - 1.22\,P_1(q) + P_0(q)\,,
\end{align*}
where $\hat{f}_1$ and $\hat{f}_2$ are the fitted models for the simulations presented in Figures~\ref{fig:Simulation1_fit} and \ref{fig:Simulation2_fit}, respectively.
By gathering terms of the same power in $q$, the derived MRDs become
\begin{align*}
&\hat{f}_1(q) =   85.6\,q^4 - 178.2\,q^3 + 129.8\,q^2 - 37.7\,q + 4.0\,,\\
&\hat{f}_2(q) =  -45.7\,q^4 + 111.0\,q^3 - 87.4\,q^2  + 21.6\,q  + 0.7\,.
\end{align*}
Differences between the MRDs derived by the two methods are found to be $\lesssim \sigma / 5$.
%where $\hat{f}_1$ and $\hat{f}_2$ are the fitted models for the simulations presented in Figures~\ref{fig:Simulation1_fit} and \ref{fig:Simulation2_fit}, respectively.

The two examples demonstrate the power of our algorithm, as the derived MRDs are very close to the underlying functions, even though the algorithm was applied without any assumption on the shape of the MRD.

%
%---------------------------------------------------------------
%---------------------------------------------------------
%---------------------------------------------------------------
%---------------------------------------------------------

%
\section{Accounting for an observational detection threshold}
% ==========================================================
%
Samples of observed spectroscopic binaries are subjected to many observational biases. An obvious one \cite[e.g.,][]{mg92, tokovinin92} is the detection threshold---RV surveys can  identified SB systems only if their RV amplitude is large enough. 
The impact of such a selection effect becomes increasingly significant for small values of $q$, causing 
the derived $\hat{f_q}$ at small $q$ values to be underestimated.
In this section we describe our way to account for this observational bias, following the approach of \cite{mg92}.

To model this effect we assume that only (and all) binaries with RV amplitude larger than some $K_{\rm min}$ are detected. 
Therefore, for each $q$ and $m_1$ there exists the longest orbital period that can be detected:  
%
%---------------------------------------------------
\begin{equation}
 % P_{\rm max}(q,m_1, K_{\rm min}) = 
  P_{\rm max} = 
   \bigg( \frac{m_1}{m_\odot} \bigg) \bigg( \frac{K_{\rm min}}{1 \text{ km/s}} \bigg)^{-3}   \frac{q^3}{(1+q)^2} \, \, 9.625 \cdot 10^6 \text{ d}  \,,
\end{equation}
%-----------------------------------------------------------
where the orbits are assumed to be circular.
For periods shorter than $P_{\rm max}$ the detectability depends on the inclination angle and therefore only a fraction of the population of binaries are detectable. We define the detection function, $D$,  which is the fraction of detected binaries out of all systems
with identical $P$, $m_1$ and $q$.
The detection function is the probability of a system
to have an inclination such that its observed RV semi-amplitude will be larger than the detection threshold, 

%----------------------------------------------------------------------
\begin{equation}
%D(q,P,m_1, K_{\rm min}) =
D =
    \begin{cases}
    \sqrt{1 - 2.21\cdot 10^{^{-5}} \, K_{min}^{^2} \, \frac{P^{^{2/3}}}{m_1^{^{2/3}}} \, \frac{(1+q)^{4/3}}{q^2}  }  & \text{if } P<P_{\rm max} \,,
    \\
    0 & \text{else,}
    \end{cases}
\label{EQ: D1}
\end{equation}
%---------------------------------------------------------------
where $P$ is in days, $m_1$ is in solar mass and $K_{\rm min}$ is in km/s.

Let us further assume that the primary stars in the sample are of nearly identical 
mass, $\overline{m}$, 
and that the distribution of the orbital period, $f_P$, is independent of 
$q$. The fraction of detected systems with some specific $q$ is composed of the probability that both the period and the inclination allow a detection, namely
%------------------------------------------------------
\begin{equation}
\overline{D}(q) = \int_{P_1}^{P_2}{  {D}(q,P,\overline{m}, K_{\rm min}) \, f_P(P)\,dP}\,,
\label{EQ: D2}
\end{equation}
%-------------------------------------------------------
%
where $P_1$ and $P_2$ are the shortest and longest periods of the population, respectively.

The derived $\hat{f_q}$ can now be corrected by the detection function, namely
%
%----------------------------------------------------------------
\begin{equation}
\hat{h}_q(q) = \frac{\hat{f_q}(q)}{\overline{D}(q)} \,,
\end{equation}
%-----------------------------------------------------------------
%
where $\hat{h}_{q}(q)$ is the unbiased distribution of $q$. This time the corrected function has to be normalized in order to be used as a PDF.  
In the case of a boxcar fit, the correction factor of each bin is taken according to its value at the bin's center.
Notably, for very small mass ratios 
the correction factor, $1/\overline{D}(q)$, becomes very large and consequently uncertain. It is therefore advised to cautiously address only a domain where the correction factor is a small number, say, $1/\overline{D}(q) \lesssim 2$.

%
%------------------------------------------------------------------
\begin{figure*}
	\centering
	\includegraphics[width=16cm]{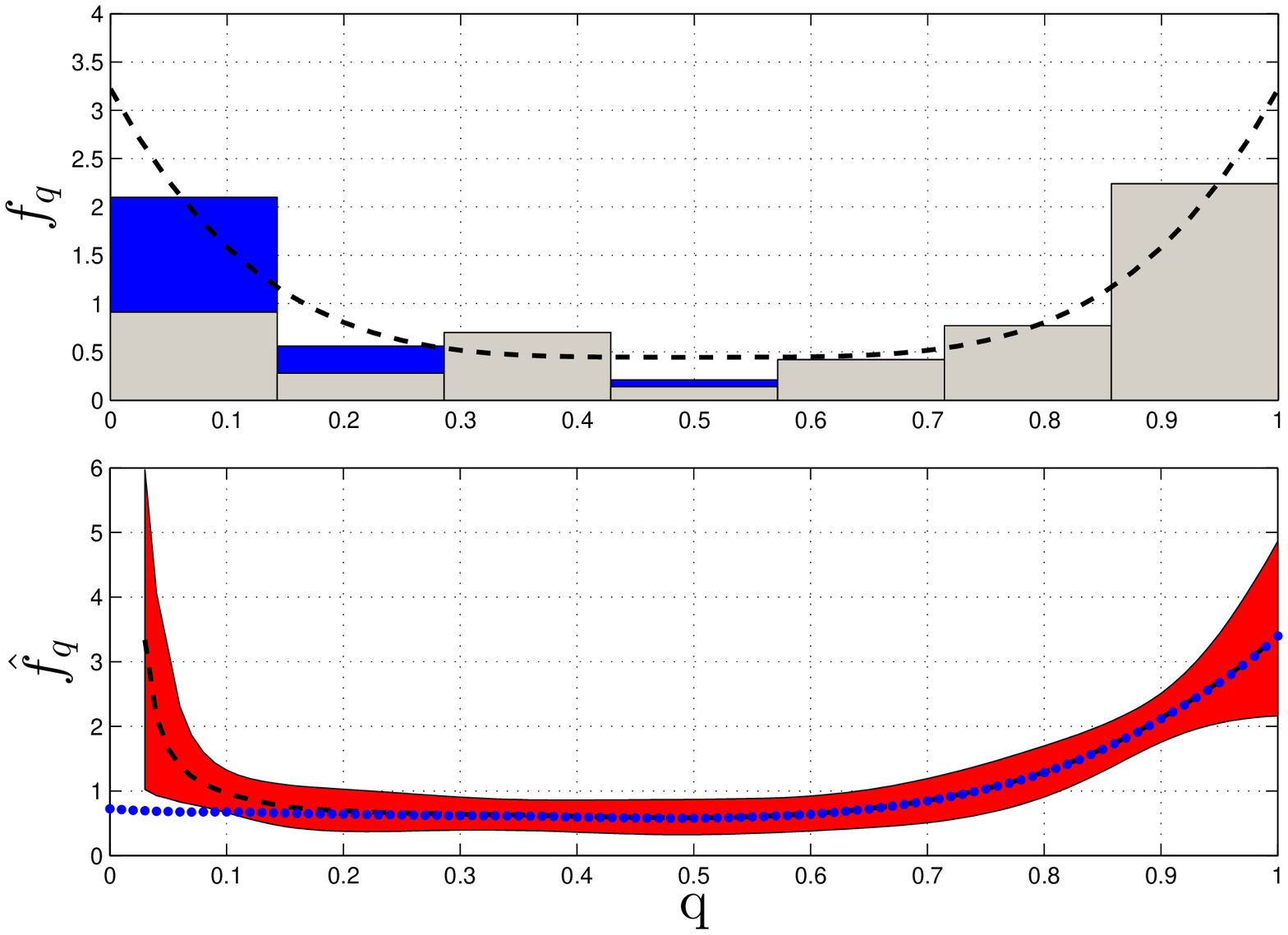}
	\caption{MRD Derivation from a simulated SB1 sample with a detection threshold of $K_{\rm min}=3$ km/s.
Top: Simulated sample of $100$ SB1s with random orientations, using an MRD (dashed black line) that peaks at $q=0$ and $q=1$. 
The specific drawn sample is presented by a seven (gray) bin histogram (see 
Figure~\ref*{fig:Simulation3_fit}). Because of the detection threshold, $22$ systems (light blue) upper bins, were not detected. 
Bottom: Derived MRD (dotted line), using the first {\it five} shifted Legendre polynomials as a basis, together with the corrected MRD (dashed line) and its confidence $1\sigma$ range (see text).
}
	\label{fig:Simulation3_fit}
\end{figure*}
%------------------------------------------------------------------

Again, to test the correcting part of the algorithm we ran numerous simulations, one of which is presented in Figure~\ref{fig:Simulation3_fit}. We used here the same population as in 
Figure~\ref{fig:Simulation1_fit}, but now  with $1M_{\odot}$ primary for each binary, orbital periods with log-uniform distribution between $1$ to $10^3$ days, and a detection threshold of $K_{\rm min}=3$ km/s, which caused 22 simulated binaries not to be detected. A histogram of the detected and missed binaries is plotted in the top panel of Figure~\ref{fig:Simulation3_fit}. As can be
seen in the figure, most of the missed binaries had low mass ratio, as expected. The lower panel shows the uncorrected and corrected distributions. 
The uncorrected MRD suffers from serious suppression of its lower part, while the correction succeeded to produce the correct underlying MRD.
As expected, for small mass ratios, $q\lesssim0.05$,
the correction factor ($1/\overline{D}(q)$) became large, and  therefore, we refrained from obtaining the corrected function for this range of $q$'s.

Another way to correct for the undetected binaries, not presented here, is to apply the derived 
$\overline{D}(q)$ factor to the base functions, and use these corrected functions along the MCMC fitting. One then constructs the true MRD by using the uncorrected base functions with the parameters obtained with the MCMC.

\section{Conclusions}
%========================
%
We have presented here a novel {\it direct} algorithm to derive the mass-ratio distribution (MRD) of short-period binaries from an observed SB1 sample. The algorithm considers a parameterized family of MRDs  and finds the set of parameters that best fits the observed sample.

The algorithm consists of four parts. First, we define a new observable, the modified mass function, $S$, derived for each SB1 in the sample. We show that the distribution of the modified mass function of an SB1 sample follows the shape of the underlying MRD, turning the use of the modified mass function more advantageous than the previously used mass function, reduced mass function or the reduced mass function logarithm.
Second, given an assumed MRD, we derive the likelihood of obtaining the observed sample of SB1s with the derived modified mass functions. Maximizing this likelihood by an MCMC search enables the algorithm to find the best parameters of the underlying MRD.
Third, we  suggest to express the unknown MRD
by a basis of functions with some unknown coefficients that linearly span the space of possible MRDs. We suggest two such bases.
Fourth, we have shown how to account for the undetected systems that have an RV amplitude below a certain threshold. 
The correction is calculated per mass ratio and therefore can be applied to the derived MRD. 
Without the correction, this observational bias suppresses the derived MRD for low mass ratios. 
Numerous simulations show that the algorithm works with either of the two bases.

The algorithm is based on three simplifications. We consider here only circular orbits, we ignore the double-lined binaries, and we assume there are no uncertainties associated with the $y$'s and therefore with the $S$'s. With the present layout, it is straightforward to generalize the algorithm to include eccentric orbits and uncertainties in the $S$'s. On the other hand, ignoring the extra information about the known mass ratio of the SB2s 
\citep[e.g.,][]{mazeh03,prato07,fernandez17} 
is an obvious drawback.   A further development of the algorithm to use the SB2 information is planned for a further publication. At present, the algorithm treats those systems as SB1s.
 
The detection threshold correction presented here depends on the orbital period distribution of the binary population and on the assumption that the MRD does not depend on the binary period \citep[see discussion by][which put this assumption into  question for O- and B-type stars]{moe17}. These two assumptions are inherent to any correction algorithm, as the RV amplitude does depend on the mass ratio and the orbital period. The simulations showed that  the correction succeeded to produce the correct MRD for low mass ratios.

The correction is based on a simplistic conception of the detection threshold. In reality, the observational bias does not act as a stiff threshold, but instead the detection probability of a binary is a continuous monotonic increasing function of its amplitude, which depends on the period, determined by the time stamp of the RV observations. However, it is quite easy to adopt the algorithm to any detection sensitivity through equations~({\ref{EQ: D1}) and (\ref{EQ: D2}), by which one can derive a more sophisticated correction for any value of mass ratio. Needless to say, any derivation of the mass ratio distribution can be based only a sample that was obtained by a complete systematic survey that searches for spectroscopic binaries with known detection thresholds, so that the corrections can be derived and applied to the observed sample.

Obviously, the correction procedure introduces additional errors to the derived MRD, due to an inexact period distribution and inaccurate detection threshold used. Therefore, the correction becomes less valuable for low mass ratios, a range for which we have small number of systems and the correction factor becomes large. In the simulated case presented above, for example, we refrained from plotting the corrected MRD for mass ratio smaller than $0.05$. The exact limit depends on the specific SB1 sample. 

In the next paper of this series (Shahaf et al., in preparation) we apply the algorithm to a few samples published in the literature, in particular those of \cite{mazeh03}, \cite{fisher05}, \cite{prato07}, \cite{mermilliod07} \citep[see also][]{north14,swaelmen17} and \cite{tal-or15}. 

Furthermore, we anticipate in the near future extremely large new samples of SBs coming from the APOGEE\footnote{http://www.sdss3.org/surveys/apogee.php} 
and the Gaia\footnote{http://sci.esa.int/gaia/}
projects. The release of the Gaia distances will enable us to better estimate the primary masses of these samples, a key element in the derivation of the reduced and modified mass function. The new algorithm will be ready for these large samples to determine the MRD of spectroscopic binaries. In addition, we anticipate two additional large samples---eclipsing binaries from large photometric data bases \citep[see, for example][for the analysis of the OGLE LMC binaries]{mazeh06,mowlavi17}, and astrometric binaries from Gaia, exploring the binaries with very short and very long period range. The new samples will finally give us the full picture of the different binary populations.

\section*{Acknowledgments}
%====================
%
We are indebted to the referee for the thorough reading of the manuscript and very helpful comments. We thank Shay Zucker for the insightful discussions.
We acknowledge support from the Israel Science
Foundation (grant No.~1423/11) and the Israeli Centers of
Research Excellence (I-CORE, grant No.~1829/12).

\bibliographystyle{mnras}
\bibliography{MR_bib}

%%%%%%%%%%%%%%%%% APPENDICES %%%%%%%%%%%%%%%%%%%%%
%==============================================

%
\newpage

\appendix
%====================
%
\section{Distribution of the reduced mass function}
% ==================================================
\label{APP: reduced mass function distribution}
The mass-ratio of a binary system is $q\equiv m_2/m_1 $,
where $m_1$, $m_2$ are the stellar masses of the primary and secondary, respectively.
The reduced mass function, $y$, is
%----------------------------------------------------------------------------
\begin{equation} \label{EQ: norm mass function APP}
	y  = \frac{q^3}{(1+q)^2} \,  {\rm sin}^3i \,.
\end{equation}
%-------------------------------------------------------------------------
%
where $i$ is the inclination. Notably, each value of $y$ is associated with a minimal possible $q$ value, that can be determined by setting the inclination angle $i$ to be $90^{\circ}$. 
This $q$ minimum, denoted $\qmin$, is the only real root of the polynomial $P_{y}(q)$,
\begin{equation} 
		P_{y}(\qmin)=y^{-1}\qmin^3 -\qmin^2 - 2\qmin -1 = 0\,.
\end{equation}
The explicit expression for $\qmin$, as was given by \cite{heacox95}, is
\begin{equation} \label{EQ: qmin}
\qmin = h(y) + \frac{1}{h(y)} \, \bigg(\,\frac{2}{3}y + \frac{1}{9}y^2\,\bigg) \, + \frac{1}{3}y  \,,
\end{equation}
where
%--------------------------------------
\begin{equation}
h(y)=\bigg( \frac{1}{2}y +\frac{1}{3}y^2 + \frac{1}{27}y^3 +\frac{\sqrt{3}y}{18}\sqrt{(4y+27)}   \bigg)^{1/3}  \,.
\end{equation}
%--------------------------------------

A rigorous development of the $y$ probability density function (PDF), $f_y$, for a sample of randomly oriented binaries has been previously presented by \cite{heacox95}.
Nevertheless, an alternative geometrical derivation of it may be instructive in the context of this work.

We choose to work in the parameter plane of $(1-{\rm cos} \, i , q)$, where $0\leq q \leq 1$ and $0 \leq 1 - {\rm cos} \, i \leq 1$, as the distribution of $1-\cos i$ is uniform for random orientation of the orbits. 
Equation~(\ref{EQ: norm mass function APP}) implies that $y$ values are uniquely associated with contours on that plane, as demonstrated in Figure \ref{fig:y_contours}. 
A specific system with some given $y\pm {\delta y}/2$ and $q \pm {\delta q}/2$ 
occupies an area on the parameter plane, 
%
%--------------------------------------------------------------------------------
\begin{equation}
\delta A = 
\delta (1-cos\,i) \, \delta q 
\approx  \bigg|\, \frac{\partial cos \, i}{\partial y} \,\bigg|\,  \delta y \, \delta q \,,
%\equiv \mathbb{K}(y,q) \, \delta q \, \delta y  \ .
\end{equation}
%______________________________________________________________
%
where by means of equation~(\ref{EQ: norm mass function APP}), $\big|\, \frac{\partial cos \, i}{\partial y} \,\big|$ is
\begin{equation}
\mathbb{K}(y,q) \equiv  \bigg|\, \frac{\partial cos \, i}{\partial y} \,\bigg| = \frac{(1+q)^{4/3}}{ 3 \, y^{1/3} \, q \,  \sqrt{ q^2 - y^{2/3}(1+q)^{4/3} }  }  \,.
\label{EQ: K(y,q)}
\end{equation}	
An example of $\delta A$ assuming $0.0100 \leq y \leq 0.0101$, at $q=0.3$, is given in Figure~\ref{fig:y_contours}.
%and $0.2994 \leq q \leq 0.3006$ is given in Figure~\ref{fig:y_contours}.

\newpage
Since $1-{\rm cos}\, i$ is uniformly distributed, 
the probability to draw a system with specific $y$ and $q$ values from a sample of randomly oriented binaries is $\sim f_q(q) \, \delta A$,
where $f_q$ is the sample's underlying MRD.
Considering all possible values of $q$ , taking $\delta A$ to be infinitesimal and assuming $0<q\leq 1$, $f_y$ becomes
\begin{equation}
f_{y}(y\,\,;\,\, f_q) = \int_{\qmin}^1 {f_q(q) \cdot \mathbb{K}(y,q) \,\, dq } \,. 
\label{EQ: f_y APP} 
\end{equation}
Some attempts have been made to use the PDF of $log(y)$, $f_{\log y}$, as a more informative representation of the data \cite[e.g.,][]{boffin10,boffin15}. In terms of equation~(\ref{EQ: f_y APP}),  $f_{\log y}$ is
\begin{equation}
f_{\log y}(u \,\,;\,\, f_q) \propto 10^u  \, \cdot \, f_{y}(10^u\,\,;\,\, f_q) \,.
\end{equation}
%

%----------------------------------------------------------------
\begin{figure*}
	\centering
	\includegraphics[width=0.9\linewidth]{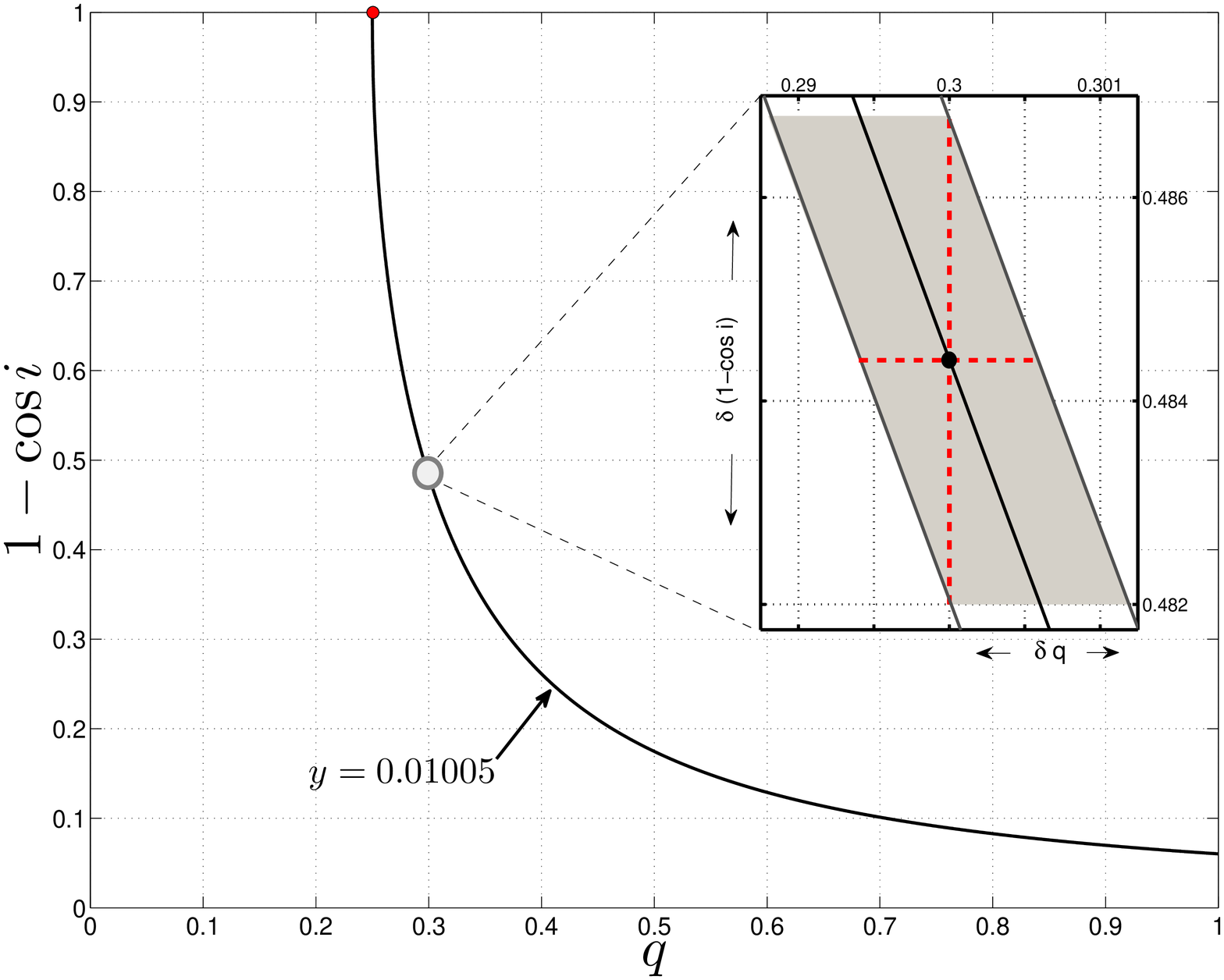}
	\caption{A contour of $y=0.01005$ plotted in the $(1-\cos i) , q)$ plane. 
	The red dot corresponds to the q minimum value of $y=0.01005$.
	Gray circle locates the point where $q=0.3$.
	The zoomed window shows the area bounded by $0.0100 \leq y \leq 0.0101$. 
	The horizontal width of the parallelogram, $\delta q$, equals to $0.0012$ (dashed red). 
	The vertical height of the parallelogram, $\delta (1-\cos i)$, equals to  $\mathbb{K}(y,q)\times \delta y = 0.0047 $ (dashed red). 
	}
	\label{fig:y_contours}
\end{figure*}
%--------------------------------------------------------------------------------------
%
% 
\section{Derivation of the modified mass function}
%================================
%
\label{APP: S derivation}
The modified mass function, $S$, is required to be a strictly increasing continuous transformation of $y$, from $[0,0.25]$ onto $[0,1]$. 
Additionally, if the underlying MRD, $f_q$, is uniform---its resulting $S$ distribution, $f_S$, is required to be uniform as well.

According to equation~(\ref{EQ: norm mass function}) ${\rm cos}\, i$  can be expressed in terms of $y$ and $q$, 
\begin{equation}
\label{EQ: cosi as func of q,y}	
{\rm cos}\, i(y,q)=\sqrt{1 - y^{2/3} \,(1+q)^{4/3} \, q^{-2}} \,.
\end{equation}
The probability to observe a system at some $y' < y$ is provided by integrating over the surface bounded by the axis, a contour of $1-{\rm cos} \, i(y,q)$ within the $(1-{\rm cos} \, i , q)$ plane, namely
\begin{equation}
P(y'<y) = \int_{A}{f_{(1-{\rm cos}i)}\cdot f_{q} \, dA} \,.
\end{equation}
Since $1-{\rm cos}\, i$ is uniformly distributed,
% simply states that $f_{(1-cosi)}=1$:
\begin{equation}
P(y'<y) = \int_{0}^{\qmin} {f_{q} \, dq} + \int_{\qmin}^{1} \big(1-{\rm cos}\, i(y,q) \big)f_{q} \, dq \,.
\label{EQ: y CDF}
\end{equation}
The modified mass function is defined by taking equation~(\ref{EQ: y CDF}) with uniform MRD:
\begin{equation}
\label{EQ:S (appendix)}
S=\mathbb{S}(y) \equiv 1 - \int_{\qmin}^{1}{\sqrt{1 - y^{2/3} \,(1+q)^{4/3} \, q^{-2}}\,dq} \,.
\end{equation}
The transformation $\mathbb{S}$ is by definition the CDF of $y$ assuming a uniform MRD, therefore it obeys the requirements given at the beginning of this subsection.

$\mathbb{S}$ is unique. Let $\mathbb{T}$ and $\mathbb{S}$ uphold the stated requirements. 
Since both are transformations of $y$, the probability density functions obey $f_S |\frac{d S}{d y}| = f_{T} |\frac{d T}{d {y}}|$. Specifically for uniform $f_q$, this relation becomes  $|\frac{d S}{d y}| = |\frac{d T}{d {y}}|$. Since both are strictly increasing and continuous  from $[0,0.25]$ onto $[0,1]$, $\mathbb{T} \equiv \mathbb{S}$.
%
%\subsection{Distribution of the modified mass function}
%----------------------------------------------------------------------
%
%Since $S$ is by definition the CDF of $y$ assuming a uniform MRD, its derivative in respect to $y$ is simply ${d S}/{d y} = f_{y}(y\,\,;\,\, f_q\equiv 1)$.
%In these terms, the $S$ PDF is
%
%\begin{equation} 
%f_S(S \,\, ; \, \, f_q) = f_{y}\big(y(S) \,\, ; \,\, f_q \big) \cdot \bigg|\frac{dy}{dS} \bigg| =
%\frac{f_{y}\big(y(S) \,\, ; \,\, f_q \big)} { f_{y}\big(y(S) \,\, ; \,\, f_q \equiv 1 \big)} \ ,
%\label{EQ: f_S in terms of f_y APP}
%\end{equation}
%
%where $y(S)$ is the inverse of $\mathbb{S}$.
%
%By plugging the explicit expression for $f_y$ from equation (\ref{EQ: f_y}) into equation (\ref{EQ: f_S in terms of f_y APP}), $f_S$ becomes
%
%\begin{equation}
%f_S(S \,\, ; \, \, f_q) = {\int_{\qminS}^1 {f_q(q) \, \mathbb{K}(y(S),q) \, dq }} \bigg/  
 %{\int_{\qminS}^1 {\mathbb{K}(y(S),q) \, dq }}  \ .
%\label{EQ: f_S}
%\end{equation}
%
%Effectively, $f_S$ is an average of $f_q$ for a given value of $y$, taken over a $y$ contour on the $( \, 1-{\rm cos} \, i \, , \, q)$ plane and weighted according to the relative area $\delta A$.
%%%%%%%%%%%%%%%%%%%%%%%%%%%%%%%%%%%%%%%%%%%%%%%%%%
%
%
%

% Don't change 
% Don't change these lines
\bsp	% typesetting comment
\label{lastpage}
\end{document}